\documentclass[
  aps,
  physrev,
  twocolumn,
  preprintnumbers,
  superscriptaddress,
  nofootinbib,
  floatfix
]{revtex4-2}

\usepackage[english]{babel}
\usepackage[autostyle,english=american]{csquotes}
\MakeOuterQuote{"}
\usepackage{float}
\usepackage{graphicx}
\usepackage{bm}
\usepackage{amsmath}
\usepackage{amssymb}
\usepackage{multirow}
\usepackage{booktabs}
\usepackage[hyperfootnotes=false]{hyperref}

\newcommand{\beq}{\begin{equation}}
\newcommand{\eeq}{\end{equation}}

\newcommand{\Korea}{Department of Physics and Astronomy, Seoul National University, Seoul 151-747, Korea}
\newcommand{\Rochester}{Department of Physics and Astronomy, University of Rochester, Rochester, NY  14627, USA}

\begin{document}

\title{Precision Measurements of the Electroweak Mixing Angle in the Region of the Z pole at the Large Hadron Collider}

\author{Arie~Bodek}
\email{bodek@pas.rochester.edu}
\affiliation{\Rochester}
\author{Hyon-San~Seo}
\email{hyon.san.seo@cern.ch}
\affiliation{\Rochester}
\author{Un-Ki~Yang}
\email{ukyang@snu.ac.kr}
\affiliation{\Korea}

\date{\today}


\begin{abstract}
In the Standard  Model (SM), the effective leptonic  weak mixing angle $\sin^2\theta^\ell_{\rm eff}$ can be  predicted to very high accuracy  by using the experimental values of  the mass of the Z boson ($M_Z$),   the Fermi constant, the fine structure constant, the  mass of  Higgs boson, the quark masses,   and  the value of the Quantum Chromodynamics (QCD)  strong coupling ($\alpha_s$) at $M_Z$. Therefore, deviations of measurements from SM predictions can be indications of new physics. With recent advances in experimental and analysis techniques, measurements  of  $\sin^2\theta^\ell_{\rm eff}$ at the Large Hadron Collider (LHC),   extracted from  the forward-backward asymmetry ($A_{\rm FB}$) in  Drell-Yan dilepton events, are now competitive with  previously published measurements at   $e^+e^-$ collider experiments. We report on  a more precise extraction of the effective leptonic weak mixing angle via an updated analysis of the CMS Drell-Yan $A_{\rm FB}$ measurements at 13 TeV.   We obtain $\sin^2\theta^\ell_{\rm eff}$=0.23154$\pm$0.00024 with additional PDF constraints from measurements of the W decay lepton charge asymmetry at 13 TeV and the W/Z cross section ratios. It is the most precise single experiment measurement to date and is in excellent agreement with the SM 2026 indirect prediction of 0.23152$\pm$0.00006. 

\end{abstract}
%


\maketitle

%
%
The weak mixing angle $\sin^2\theta_{\rm W}$ is a fundamental parameter of the Standard Model  (SM) of particle physics.  In the on-shell scheme,  $ \sin^2\theta_{\rm W}$ = 1-$M^2_W/M^2_Z$, where
$M_W$ and $M_Z$ are the masses of the  $W$  and $Z$ bosons, respectively.  The effective leptonic weak mixing angle   $\sin^2\theta^\ell_{\rm eff}$ (which can be expressed in terms of   $\sin^2\theta_{\rm W}$ with a radiative correction factor) is the quantity more directly extracted from experiment.  

 In the SM $\sin^2\theta_{\rm W}$,  $\sin^2\theta^\ell_{\rm eff}$  and  $M_W$ are  predicted to very high accuracy  by using the experimental values of of the mass of the Z boson ($M_Z$), 
  the Fermi constant, the fine structure constant, the  mass of  Higgs boson, the quark masses, 
  and  the value of the Quantum Chromodynamics (QCD)  strong coupling  ($\alpha_s$) at $M_Z$.  The  2026 SM  indirect determinations\cite{PDG2026} are
$\sin^2\theta^\ell_{\rm eff}(M_Z)$=0.23152$\pm$0.00006, and
$M_W$= 80.3570$\pm$0.0060 GeV. 
The primary motivation for more precise measurements of $\sin^2\theta^\ell_{\rm eff}$ is that comparisons between these SM predictions and precision measurements provide sensitive tests of the electroweak sector.
 
 The two most precise measurements of $M_W$
are  in disagreement.
The measurement by the CDF collaboration \cite{CDF:2022hxs} of  80.4335 $\pm$ 0.0094 GeV is about 9 standard deviations higher than the SM value, and the measurement
by the CMS collaboration\cite{CMS:2024lrd}  of   80.3602 $\pm$ 0.0099 GeV is in good agreement with the SM.  The CDF measurement of  $M_W$ can be accommodated by introducing  new physics  processes beyond the SM. For example,
using the  CDF measurement of $M_W$ as input a Two Higgs Doublet  Model\cite{Biekotter:2022abc} (2HDM) predicts  $\sin^2\theta^\ell_{\rm eff}$ = 0.23110 $\pm$ 0.00010. This provides a second motivation fo continued precision studies of $\sin^2\theta^\ell_{\rm eff}$ as a complementary electroweak observable.
%
\begin{figure}[ht]
\begin{center}
\includegraphics[width=3.in,height=1.7in]{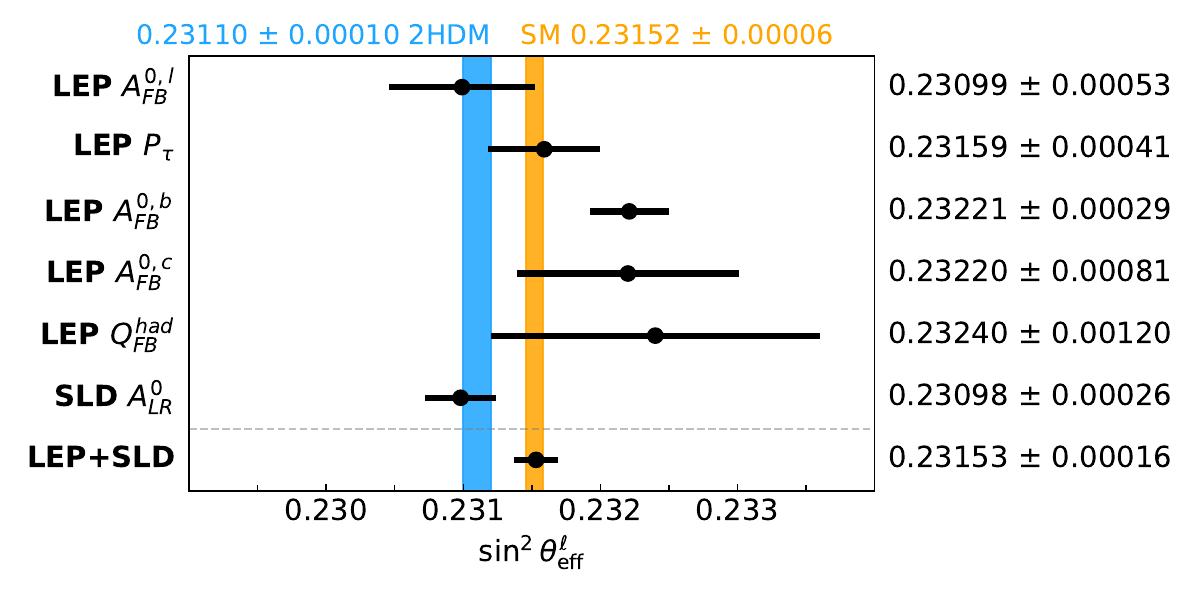}
\vspace{-0.25in}
\caption{ Measurements of  $\sin^2\theta^\ell_{\rm eff}(M_Z)$  in electron colliders (LEP, SLD)
\cite{ALEPH:2005ab} compared to
the prediction of the 2026 SM  global fit~\cite{PDG2026}.  Also shown is the prediction
of the Two Higgs Doublet Model\cite{Biekotter:2022abc} corresponding to the   CDF $M_W$ value\cite{CDF:2022hxs}
(80.4335 $\pm$0.0094  GeV).  The average  of all  six different LEP/SLD measurements of
  0.23153$\pm$0.00016  is in  agreement with the SM value (with  $\chi^2/\mathrm{ndf}$ of 11.8/5, p=0.038.)
}
\vspace{-0.3in}
\label{Fig_1}
\end{center}
\end{figure} 

The most precise extractions of 
$\sin^2\theta^\ell_{\rm eff}$ are from measurements of asymmetries  in the region of the $Z$ pole in $e^+e^-$ colliders which are shown in Fig.\ref{Fig_1}.
The average of the six LEP and SLD determinations is $0.23153\pm0.00016$, consistent with the 2026 SM indirect determination. The spread among the measurements gives $\chi^2/{\rm ndf}=11.8/5$, corresponding to a $p$ value of 0.038.
This originates from the fact that the two most precise measurements 
of $\sin^2\theta^\ell_{\rm eff}$
eff in $e^+e^-$ colliders differ by
3.2 standard deviations. A value of
0.23221±0.00029 from the measurements
of the b-quark forward-backward asymmetry $A^{0,b}_{FB}$
at the LEP experiments at
CERN, and a value of 0.23098±0.00026 from the measurement of the asymmetry of Z boson production cross sections
for left-handed and right-handed electrons ($A_{LR}$) by the SLD experiment at SLAC. It is interesting to note that the SLD
ALR measurement of $\sin^2\theta^\ell_{\rm eff}$
eff is lower than the SM prediction and is more consistent with the predictions of 2HDM,
while the LEP measurement from the b-quark forward-backward asymmetry (A0,b
FB) is higher than the SM prediction. This provides a third motivation fo continued precision studies of $\sin^2\theta^\ell_{\rm eff}$.

%
%
  With recent advances in experimental and analysis techniques\cite{Bodek:2012id,Bodek:2010qg,Bodek:2015ljm}, measurements  of  $\sin^2\theta^\ell_{\rm eff}$  at the Large Hadron Collider (LHC)   extracted from  the forward-backward asymmetry in  Drell-Yan dilepton events can be competitive with the previously published measurements at   $e^+e^-$ collider experiments.
Because the leading uncertainties in the LHC extractions differ from those in $e^+e^-$ measurements, with PDF uncertainties playing a central role at the LHC, these measurements provide complementary precision tests of the electroweak sector. 
 In this paper we report  on  the more precise extractions of $\sin^2\theta^\ell_{\rm eff}$ by an updated analysis of published asymmetries  of Drell-Yan dilepton events at the Large Hadron Collider (LHC) at 13 TeV by the CMS experiment.

The extraction is based on the polar-angle distribution of the negative lepton in the Collins-Soper frame\cite{Collins:1977iv}. 
The angle $\theta$ is measured with respect to the incoming-quark direction, which defines the forward direction for $A_{\rm FB}$. 
After integration over the $Z$ boson transverse momentum, the angular distribution can be written as
\begin{equation}
\frac{d\sigma}{d\cos\theta}\propto
1+\cos^2\theta +  \frac{A_0}{2} (1-3\cos^2\theta)  +A_4\cos\theta,
\end{equation}
where $A_0$ and $A_4$ are functions of the rapidity ($y$) and mass ($M$) of the dilepton pair.
For full acceptance, integration over $\cos\theta$ gives $A^{\rm full}_{\rm FB} = (3/8)A_4$.
In proton-proton collisions, the incoming-quark direction is not known event by event. 
It is estimated from the boost direction of the dilepton pair, because valence quarks carry, on average, more momentum than sea antiquarks.

 %
 Recently, the CMS collaboration at the LHC  reported\cite{CMS:2024ony}  on a  measurement of   $\sin^2\theta^\ell_{\rm eff}$ (0.23152$\pm$0.00031) extracted from the forward-backward asymmetry $A_{\rm FB}$ in Drell–Yan dilepton production in proton-proton collisions at a center of mass energy $\sqrt{s}=$ 13 TeV with an integrated luminosity of 138 fb$^{-1}$.  The statistical error in this measurement is $\pm$0.00010. 
  The experimental systematic error is small  because of the precise lepton momentum calibration\cite{Bodek:2012id} and the use of the angular weighted\cite{Bodek:2010qg}  
 asymmetry  $A^w_{\rm FB}$. The experimental systematic errors in acceptance and efficiencies mostly cancel in the measurement of $A^w_{\rm FB}$, which can be regarded as the detector-level equivalent of full-acceptance asymmetry.

The dominant source of uncertainty in the CMS measurement comes from uncertainties in the proton PDFs. 
For this reason, the CMS analysis compared several PDF sets from the CTEQ~\cite{Hou:2019efy,Xie:2023qbn,Hou:2022onq,CTEQ}, NNPDF~\cite{NNPDF:2021njg,NNPDF:2024djq,NNPDF:2024dpb,NNPDF:2024nan,NNPDF}, and MSHT~\cite{Bailey:2020ooq,Cridge:2021pxm,McGowan:2022nag,Cridge:2023ryv,MSHT} groups. 
The PDF uncertainties in each PDF set, as well as the differences between $\sin^2 \theta^\ell_{\rm eff}$  values extracted using different PDF sets, are  significantly reduced by using the  $A_4$ PDF re-weighting/profiling technique (first proposed in \cite{Bodek:2015ljm}) to constrain PDFs as described below. In general, PDF re-weighting/profiling is a method of further constraining the PDF parameters by including new data that has not been used in the original PDF fits.  
The \textsc{CT18Znnlo}~\cite{Hou:2019efy} set was chosen as the nominal PDF because the extracted value is near the middle of the values obtained with the other PDF sets and the set provided good coverage of the other central values. It also shows a relatively small central-value shift after profiling.

%
\begin{figure*}[t]
\begin{center}
\includegraphics[width=6.5in,height=3.5in]{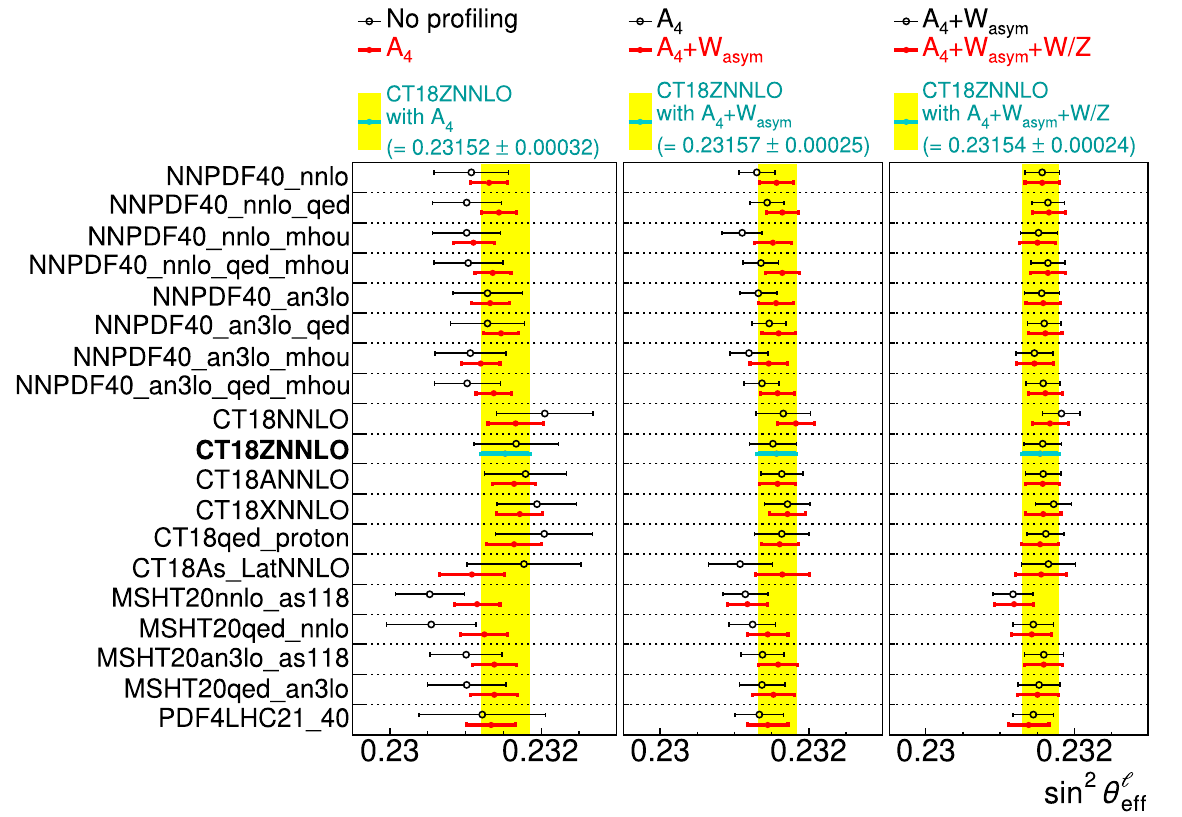}
\vspace{-0.2in}
\caption{Values of  $\sin^2\theta^\ell_{\rm eff}$  (extracted with 19 different PDF sets) before profiling compared to values after profiling with the CMS 13 TeV  $A_4$  data (left panel, 63  data points) and by also including the  
  CMS $W_{\rm asym}$  measurement\cite{CMS:2020cph} at 13 TeV  (middle panel, 18 additional data points)  and then by also including the  CMS measurement of the  $W$ and $Z$ cross section ratio\cite{CMS:2024myi} at 5.02 and 13 TeV (right panel, 2 additional points).
}
\vspace{-0.25in}
\label{Fig_2N}
\end{center}
\end{figure*}

In addition to the extraction of  $\sin^2\theta^\ell_{\rm eff}$ at 13 TeV, the CMS collaboration reported on the extraction of the mass  and rapidity dependent angular coefficient $A_4$ (= $8A_{\rm FB}/3$). The extraction  and unfolding of $A_4$
requires very good  modeling of the detector acceptance,  efficiencies and resolutions.  If the  modeling of efficiencies,  acceptance and resolutions are correct, then the  extracted value of  $\sin^2\theta^\ell_{\rm eff}$ from the unfolded $A_4$ should be  in agreement with the value extracted from the angular weighted  $A^w_{\rm FB}$.  This consistency has been achieved in the CMS analysis. The published unfolded $A_4$ values therefore provide a reusable input for future re-extractions of $\sin^2\theta^\ell_{\rm eff}$, either by adding new PDF constraints, as done in this paper, or through combinations with other (or future) hadron-collider $A_4$ measurements.

 \begin{figure*}[t]
\includegraphics[width=6.5in,height=3.in]{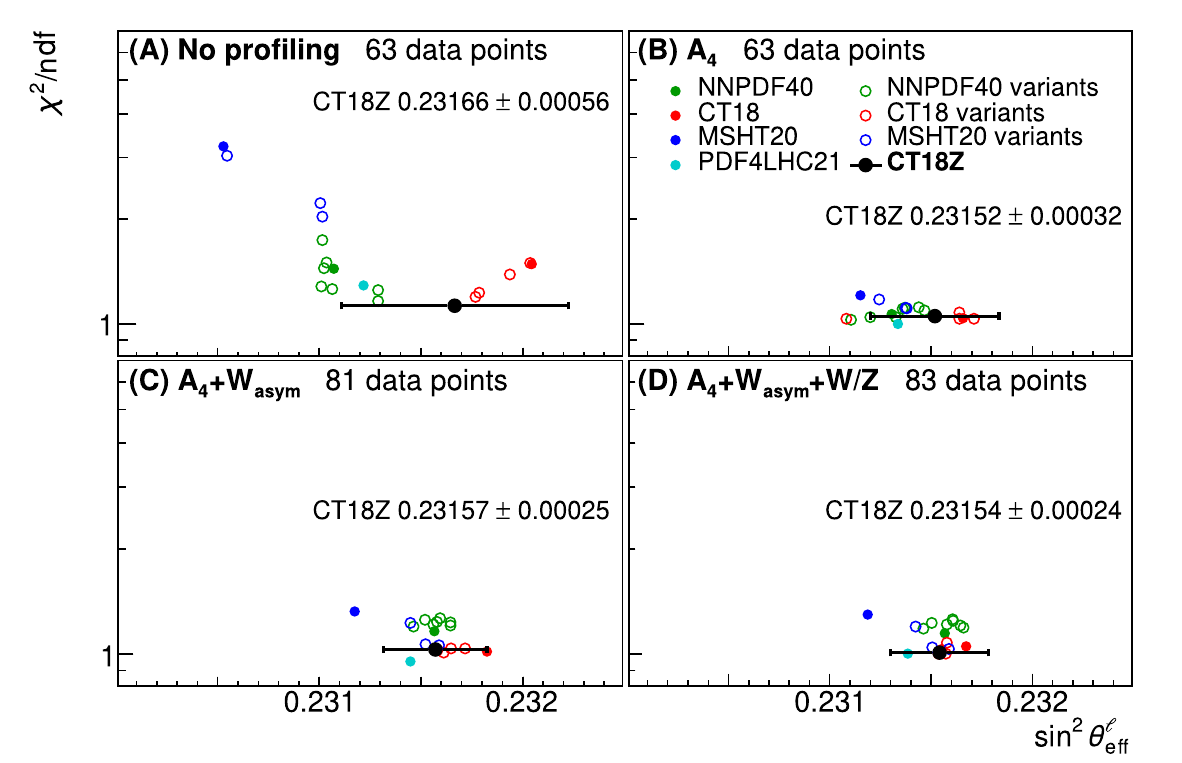}
\vspace{-0.2in}
\caption{Extracted values of $\sin^2\theta^\ell_{\rm eff}$ from the 13 TeV CMS $A_4$ data  for 19 different PDF sets on the horizontal axis. (A) Before profiling.(B) After profiling with $A_4$.  (C) After profiling with $A_4$ plus $W$ decay lepton asymmetry.   (D) After profiling with $A_4$ plus $W$ decay lepton asymmetry plus W/Z cross section ratios. The vertical axis shows the $\chi^2$ values of the fits divided by the number of degrees of freedom ($N_{\rm data}-1$), where one degree of freedom corresponds to the free parameter $\sin^2 \theta^\ell_{\rm eff}$.
}
\label{Fig_3N}
\vspace{-0.1in}
\end{figure*} 
%
 %
 The PDF re-weighting/profiling technique to constrain PDFs\cite{Bodek:2015ljm}  relies on including the  rapidity and dilepton mass dependence of $A_4$  to constrain PDFs in  the analysis.  This is because  at the region of the $Z$ pole,  $A_4$  is sensitive to the vector couplings, which depend on $\sin^2\theta^\ell_{\rm eff}$, and  also to PDFs.  At higher and lower  dilepton mass $A_4$ is  sensitive to PDFs, but much less sensitive to $\sin^2\theta^\ell_{\rm eff}$. Consequently, $A_4$  measurements in the higher and lower mass  regions provide additional constraints on PDFs.


%
Here we present an updated analysis based on the published CMS measurement of $A_4$ at 13 TeV. As a starting point, we first reproduce the original CMS analysis, where the dilepton rapidity and mass dependence of $A_4$ was used to profile PDFs and extract $\sin^2\theta^\ell_{\rm eff}$ with the open-source code \textsc{xFitter}\cite{Alekhin:2014irh,HERAFitterdevelopersTeam:2015cre}, as done by CMS. 
Then, we extend this approach by including additional recent CMS measurements of $W$ and $Z$ boson production in the profiling. 
For this extended analysis, we adopt the \textsc{CT18Znnlo} PDF set as the nominal PDF, consistent with the CMS analysis, and study an extended sample of 19 other modern PDF sets.

%
%

As done  in  the CMS analysis\cite{CMS:2024ony} the value of $\sin^2\theta^\ell_{\rm eff}$ is determined in a profiling analysis, by minimizing the $\chi^2$ function
\begingroup
\begin{multline}
\chi^2(\beta_{\rm exp},\beta_{\rm th}) = \\
\sum_i
\frac{
\left(
\sigma_i^{\rm exp}
+\sum_j\Gamma_{ij}^{\rm exp}\beta_{j,{\rm exp}}
-\sigma_i^{\rm th}
-\sum_k\Gamma_{ik}^{\rm th}\beta_{k,{\rm th}}
\right)^2
}{\Delta_i^2} \\
+\sum_j\beta^2_{j,{\rm exp}}
+\sum_k\beta^2_{k,{\rm th}} .
\end{multline}
\endgroup

%




%

 \begin{figure*}[t]
\includegraphics[width=3.25in,height=3.in]{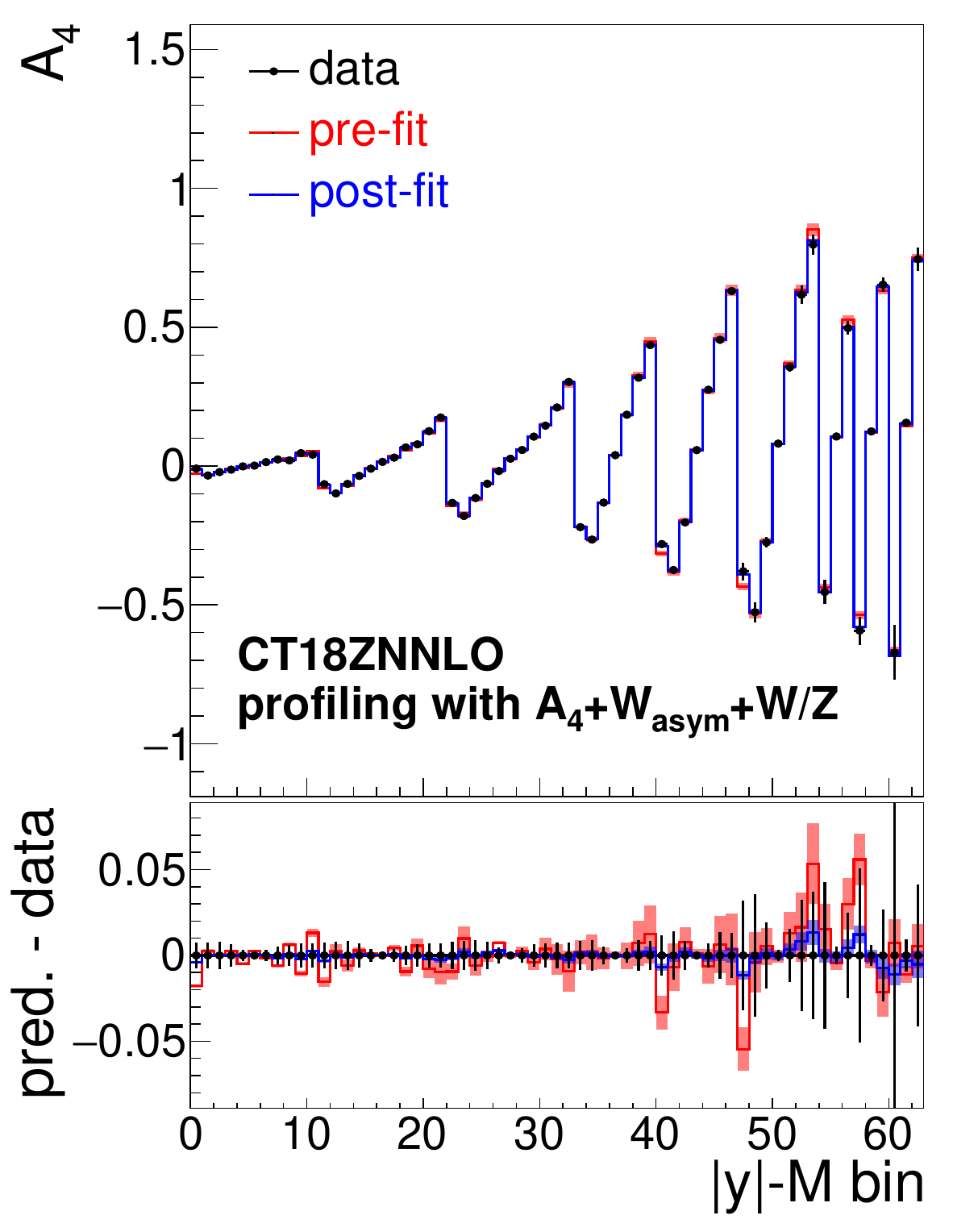}
\includegraphics[width=3.25in,height=3.in]{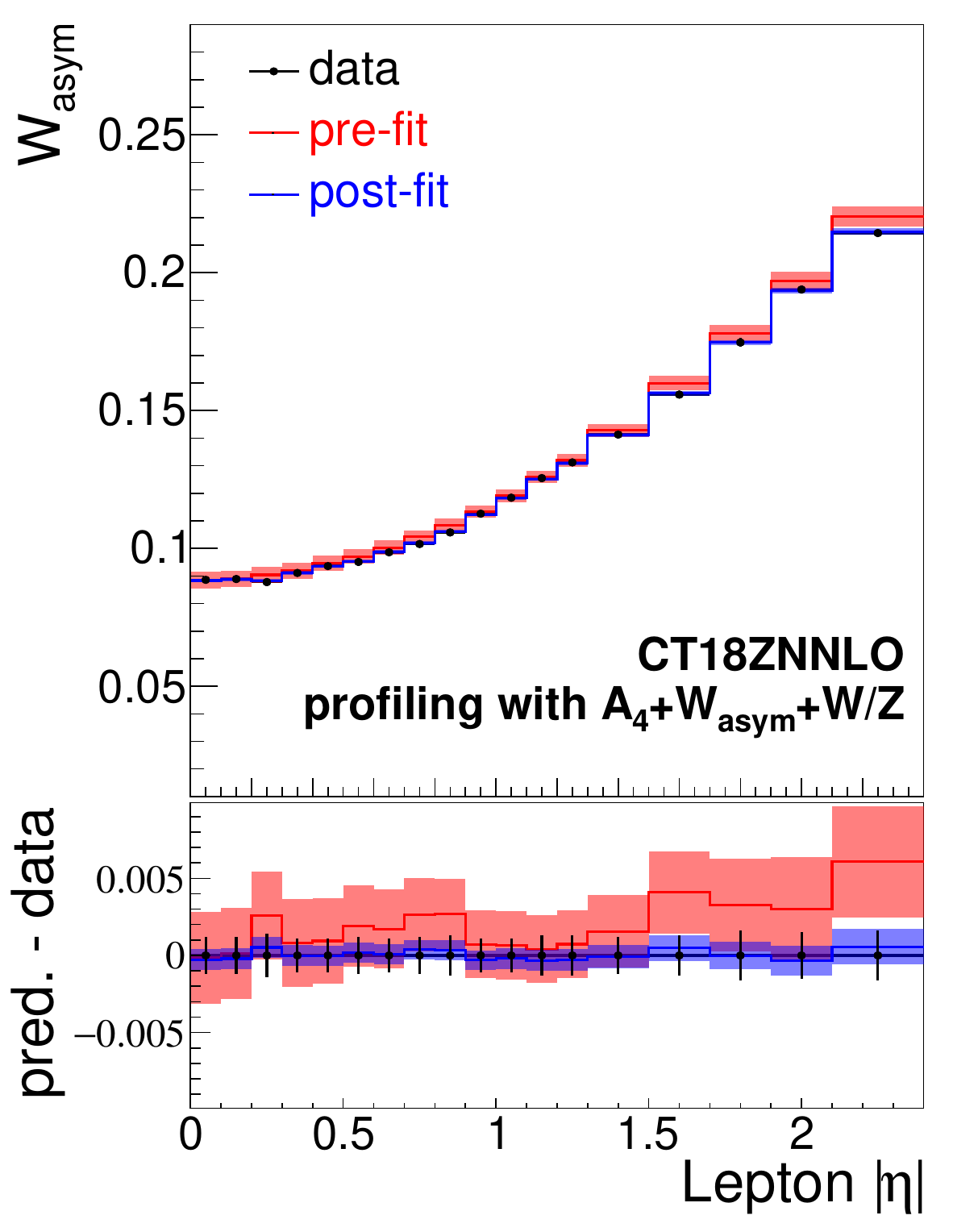}
\vspace{-0.15in}
\caption{Comparison of  predictions of \textsc{CT18Znnlo} for $A_4(|y|,M)$, and $W_{\rm asym}(|\eta|)$ 
before (red) and after (blue)  full profiling.}
%
\label{Fig_4}
\vspace{-0.15in}
\end{figure*}

The correlated experimental and theoretical uncertainties are included in the nuisance vectors $\beta_{\rm exp}$ and $\beta_{\rm th}$.
Their effects on the measured distributions and theory predictions are described by the matrices $\Gamma^{\rm exp}$ and $\Gamma^{\rm th}$.
The index $i$ runs over all $N_{\rm data}$  data points (63 for $A_4$ profiling) of the ($|y|$-$M$) double-differential $A_4$ measurement,
whereas the $j$ and $k$ indices correspond to the experimental and theoretical uncertainty nuisance parameters, respectively.
The measurements and the uncorrelated experimental uncertainties are represented 
by $\sigma_i^{\rm exp}$ and $\Delta_i$, respectively, 
whereas the theoretical predictions are denoted by $\sigma_i^{\rm th}$.
The information in the experimental covariance matrix of the double-differential $A_4$ measurement
is included in the $\Gamma^{\rm exp}$ matrix.

The theoretical prediction follows the original CMS analysis. It is obtained with the $\textsc{POWHEG}$ 2 event generator at multi-scale improved next-to-next-to-leading order ($\textsc{MiNNLO}_{\rm PS}$) accuracy in QCD~\cite{Monni:2020nks}, matched to $\textsc{PYTHIA}$ 8.2 for parton showering. Photon final-state radiation is simulated with the $\textsc{PHOTOS}$ 2 package~\cite{Golonka:2005pn}. In addition, next-to-leading order virtual weak corrections are included using $\textsc{POWHEG-Z\_{ew}}$~\cite{Chiesa:2019nqb,Chiesa:2024qzd,Barze:2013fru}.

The theoretical uncertainty includes contributions from the missing higher-order QCD and Electroweak (EW) corrections, 
as evaluated with $\textsc{ POWHEG-Z\_{ew}}~$  and PDF uncertainties by using grids generated at NLO with 
 \textsc{MadGraph5-aMC@nlo}~\cite{Alwall:2014hca} and \textsc{PineAPPL}~\cite{schwan2024nnpdf, Carrazza:2020gss}.
The matrix $\Gamma^{\rm th}$ includes nuisance parameters  reflecting missing higher-order EW corrections, the PDF Hessian uncertainties, 
and $\sin^2\theta^\ell_{\rm eff}$, which is left free in the fit. The impact of missing higher-order QCD corrections is estimated by repeating the fit for each of the six scale variations, and the maximum deviation from the central result is taken as the uncertainty.

\begin{figure*}[t]
\includegraphics[width=6.5in,height=3.in]{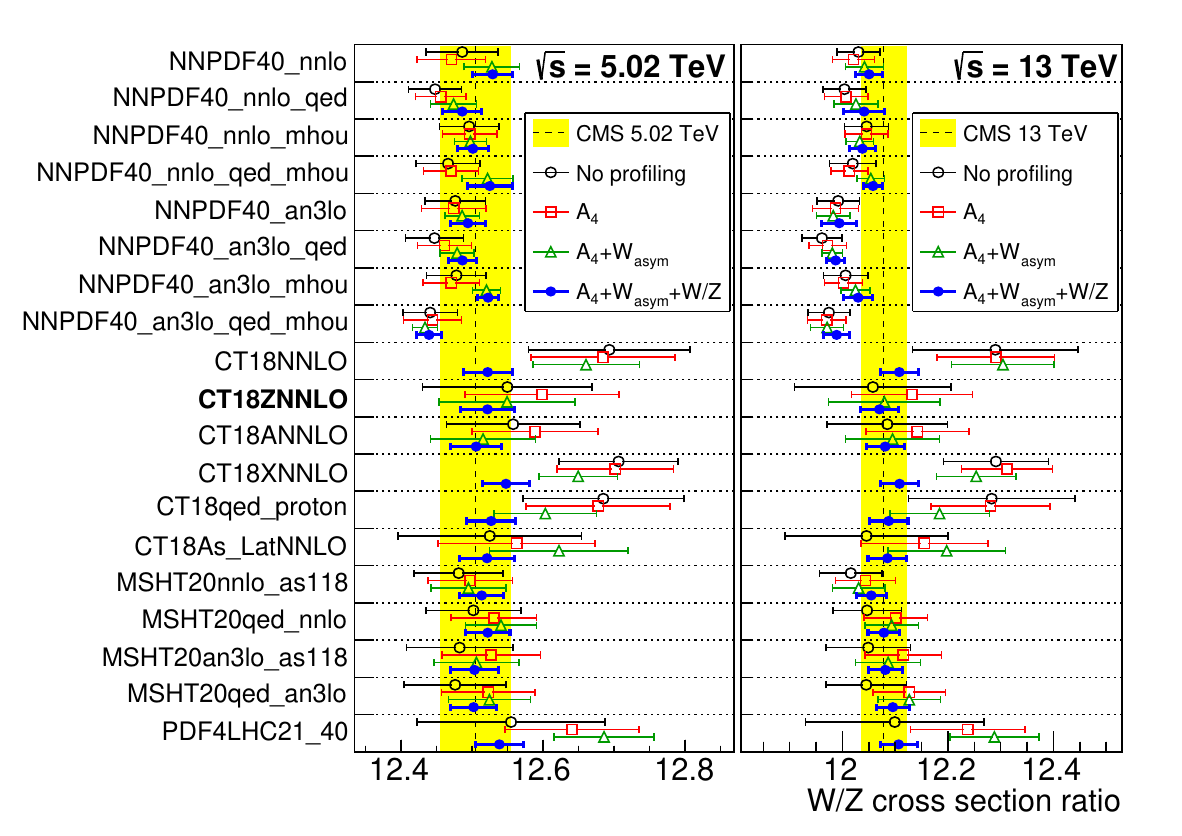}
\vspace{-0.2in}
\caption{Comparison of  the predictions of 19 PDF sets with various levels of profiling to the  W/Z fiducial cross section ratios measured by CMS at 5.02 TeV and 13 TeV}
\label{Fig_5}
\vspace{-0.1in}
\end{figure*}  

We repeat the extraction for 19 modern PDF sets, including PDFs from the CTEQ, NNPDF, and MSHT groups, as well as PDF4LHC21\_40~\cite{PDF4LHCWorkingGroup:2022cjn}. For each PDF set, values and uncertainties of $\sin^2\theta^\ell_{\rm eff}$ are extracted before and after profiling with the 13 TeV $A_4$ distribution, as shown in the left panel of Fig.~\ref{Fig_2N}. Tables of the numerical values and uncertainties are provided as Supplemental Material.
 Most of the PDFs are Next-to-Next-to-Leading-Order QCD (NNLO),
 and some are Approximate-Next-to-Next-to-Next-to LO (an3lo).  PDF sets labeled "qed" account for  QED (Quantum Electrodynamics) effects and include
photon distributions.  The NNPDF40 label “mhou” indicates that these PDF sets include corrections for  “missing higher-order uncertainties”.
The MSHT and  NNPDF PDF sets  assume  asymmetric sea strange quark densities.  The CTEQ PDF sets (except for CT18As PDF) assume symmetric strange quark densities.

 The  total uncertainties
 include contributions from  statistical, experimental systematic, theoretical, and PDF sources.  
The differences in the results obtained with the various PDF sets ({\it before profiling}) 
are mostly attributed to differences  in the  choices of experimental data, parameterizations and assumptions on the flavor decomposition.
If there are sufficient parameters in the PDF sets, then  the profiled  PDFs should be able to describe all data  used in the   $\sin^2\theta^\ell_{\rm eff}$  profiling analysis with good $\chi^2$.  
%

%
%
The uncertainty in $\sin^2\theta^\ell_{\rm eff}$ is significantly reduced after profiling with $A_4$, but some differences between PDF sets remain. As shown in \cite{Bodek:2015ljm}, including the $W$ boson decay lepton asymmetry measurements ($W_{\rm asym}$) in the profiling provides additional constraints on the $d/u$ ratio, leading to a further reduction of the PDF uncertainty in $\sin^2\theta^\ell_{\rm eff}$. Moreover, some CT18 PDF sets lack sufficient constraints on the strange-quark distribution at high energy scales, resulting in systematically lower strange-quark densities compared to other PDFs and thereby predicting a larger effective weak mixing angle. Incorporating the $W/Z$ cross section ratios provides further constraints on the strange-quark distribution and reduces these discrepancies.
   
Therefore, we  perform an updated analysis of  the published CMS 13 TeV $A_4$  data (63  data points) and also include  the  recent   
  CMS $W_{\rm asym}$  measurement\cite{CMS:2020cph} at 13 TeV  (18 additional  data points) as well as the  CMS measurement of the  
  $W$ and $Z$ fiducial cross section ratios\cite{CMS:2024myi} at 5.02 TeV ($12.505 \pm 0.037_{stat} \pm 0.032_{syst}$) and 13 TeV ($12.078 \pm 0.028_{stat} \pm 0.032_{syst}$)(2  additional data points).  
  The results are shown 
  in Figures  \ref{Fig_2N}, \ref{Fig_3N}, \ref{Fig_4}, and \ref{Fig_5}.
  
     Fig. \ref{Fig_2N} shows values of  $\sin^2\theta^\ell_{\rm eff}$  (extracted with 19 different PDF sets) before profiling compared to values after profiling with the CMS 13 TeV  $A_4$  data (left panel), by also including the 
  CMS $W_{\rm asym}$  measurement\cite{CMS:2020cph} at 13 TeV  (middle panel),  and by also including the  CMS measurement of the  $W$/$Z$ cross section ratios ( right panel).

  Extracted values of $\sin^2\theta^\ell_{\rm eff}$ from the 13 TeV CMS $A_4$ data  for different PDF sets are shown on the horizontal axis of  each of  the four panels in  Fig. \ref{Fig_3N}.  The vertical  axis shows the values of $\chi ^2$ normalized to the number of degrees of freedom, $\rm{ndf}=N_{\rm data}-1$, where the subtraction accounts for the single free parameter $\sin^2\theta^\ell_{\rm eff}$.  Shown are
the values before profiling (panel A, 63 data points), after profiling with $A_4$ (panel B, 63 data  points), after profiling with $A_4$ and also with  $W$ decay lepton asymmetry (panel C,  81  data points), and after profiling with $A_4$ and the $W$ decay lepton asymmetry and also the W/Z cross section ratios (panel D, 83 data points).  

We find that the PDF parametrization is sufficiently flexible in the relevant kinematic region, $Q\simeq M_Z$ and $x=0.001$--0.1. Our profiling is restricted to asymmetry and cross-section ratio measurements, whose uncertainties are statistically dominated. This restriction is useful because statistical uncertainties are determined directly from the data samples, whereas systematic uncertainties may depend on modeling choices and analysis-dependent assumptions. Consequently, the resulting PDF constraints are less sensitive to possible underestimates of experimental systematic uncertainties.

As seen in Figures \ref{Fig_2N} and \ref{Fig_3N}  after $A_4$ profiling  the error in the  CMS extracted value of $\sin^2\theta^\ell_{\rm eff}$ with  the nominal \textsc{CT18Znnlo} PDF set is reduced from 0.00056  to 0.00032 (in which the PDF contribution is 0.00027). By also  including the $W$ decay lepton asymmetry and the $W/Z$ cross section ratios the uncertainty for  the nominal \textsc{CT18Znnlo} is reduced to  0.00024 (in which the PDF contribution is 0.00017).

     \begin{figure}
\includegraphics[width=3.5in,height=2.7in]{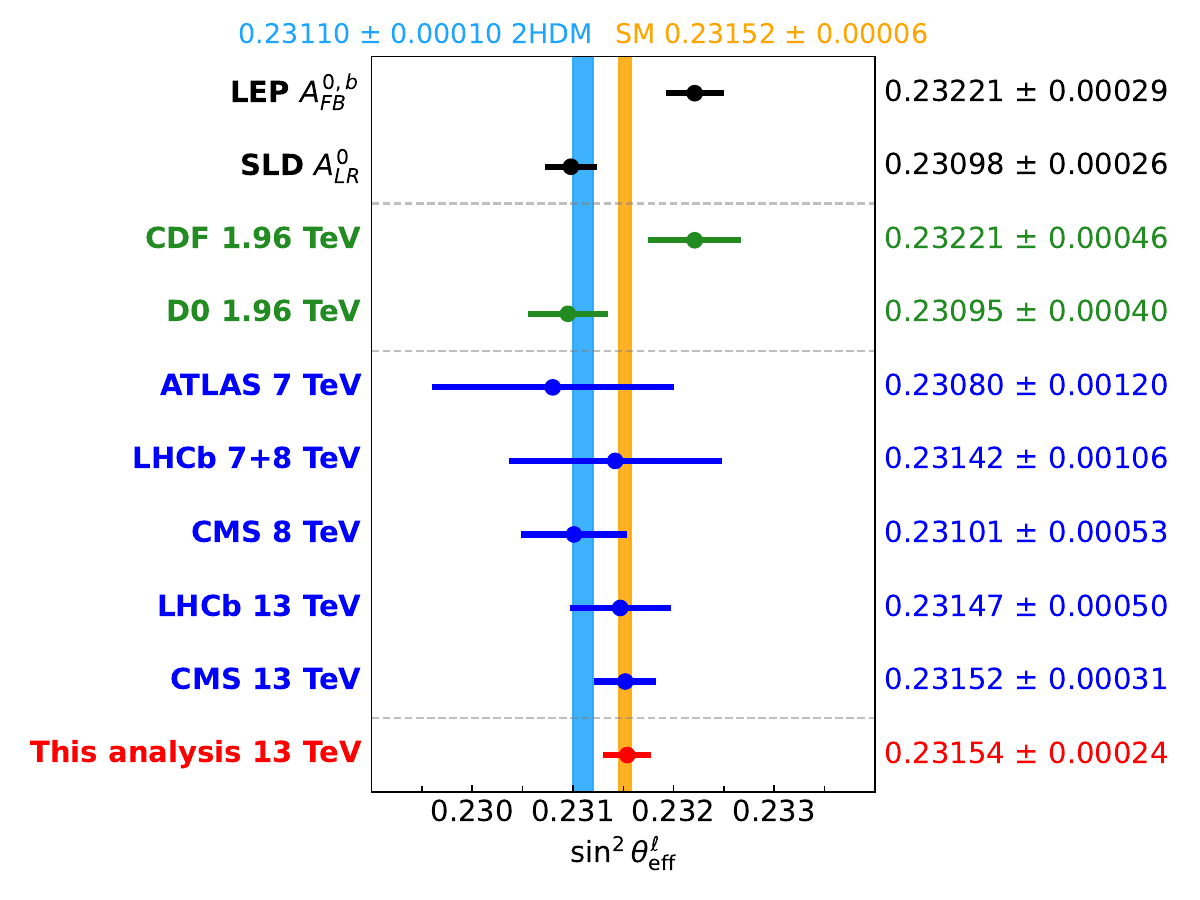}
\vspace{-0.3in}
\caption{ Comparison of the $\sin^2\theta^\ell_{\rm eff}$ value extracted this analysis (labeled "This analysis 13 TeV")
with previous measurements~\cite{ALEPH:2005ab,CDF:2016cei,D0:2017ekd,ATLAS:2015ihy,CMS:2018ktx,CMS:2024ony, LHCb:2015jyu,LHCb:2024ygc} 
and the 2026 SM indirect prediction~\cite{PDG2026}.  Also shown is the prediction
of the Two Higgs Doublet Model\cite{Biekotter:2022abc} corresponding to the   CDF $M_W$ value\cite{CDF:2022hxs} 
(80.4335 $\pm$0.0094  GeV).
}
\label{Fig_6}
\vspace{-0.25in}
 \end{figure}

After profiling with the CMS $A_4$, $W_{\rm asym}$, and $W/Z$ ratio measurements, the extracted values of $\sin^2\theta^\ell_{\rm eff}$ for 18 PDF sets are within half standard deviation of the nominal \textsc{CT18Znnlo} result. The exception is $MSHT20nnlo\_as118$, whose central value is lower than the \textsc{CT18Znnlo} value by 1.46 times the nominal \textsc{CT18Znnlo} uncertainty and whose fit gives the largest $\chi^2$ among the profiled PDF sets. The QED and approximate N$^3$LO MSHT20 variants give values closer to the nominal result and lower $\chi^2$ values, which may reflect their modified treatment of PDF, photon-PDF, and theoretical uncertainties. 

The same profiling also reduces the sensitivity to assumptions about the strange-quark distribution. After the $W_{\rm asym}$ and $W/Z$ ratio measurements are included, the extracted values of $\sin^2\theta^\ell_{\rm eff}$ show no significant difference between PDF sets with symmetric and asymmetric strange-quark distributions.

%

The quality of the profiled description of the input measurements is shown in Figs.~\ref{Fig_4} and \ref{Fig_5}. After the full profiling, the \textsc{CT18Znnlo} predictions are in good agreement with the measured $A_4(|y|,M)$ and $W_{\rm asym}(|\eta|)$ distributions. The corresponding comparison with the CMS $W/Z$ cross section ratios is shown for the 19 PDF sets and different levels of profiling in Fig.~\ref{Fig_5}.

At present, the CMS measurements used in the profiling provide more precise constraints on the relevant PDFs than the corresponding measurements available from LHCb and ATLAS. Consequently, the PDFs profiled with the CMS data can be also used to reduce the PDF uncertainties in $\sin^2\theta^\ell_{\rm eff}$ measurements of LHCb and ATLAS. The profiled PDF paarameters are given in Supplemental Material.

For the published LHCb 13 TeV measurement~\cite{LHCb:2024ygc}, $\sin^2\theta^\ell_{\rm eff}=0.23147\pm0.00050$, the PDF component of the model uncertainty is 0.00021. Using the CMS-profiled \textsc{CT18Znnlo} PDF, this PDF uncertainty is reduced to 0.00015, yielding a total uncertainty of 0.00046. For the ATLAS 8 TeV $A_4$ analysis currently under study~\cite{ATLAS:2018gqq}, the expected total uncertainty is 0.00036, with a PDF contribution of 0.00024. We estimate that using the CMS-profiled \textsc{CT18Znnlo} PDF would reduce the PDF contribution to 0.00015 and the total uncertainty to 0.00031.
 
%
Fig.~\ref{Fig_6} shows a comparison of the $\sin^2\theta^\ell_{\rm eff}$ value extracted in this analysis from the CMS 13 TeV data using the nominal \textsc{CT18Znnlo} PDF set with previous measurements~\cite{ALEPH:2005ab,CDF:2016cei,D0:2017ekd,ATLAS:2015ihy,CMS:2018ktx,CMS:2024ony,LHCb:2015jyu,LHCb:2024ygc}.
 Also shown are the 2026  SM indirect prediction~\cite{PDG2026} and the prediction of the  Two Higgs Doublet Model\cite{Biekotter:2022abc}
   corresponding to the CDF $M_W$ value\cite{CDF:2022hxs} of 80.4335 $\pm$0.0094  GeV.



In conclusion, by including the CMS 13 TeV $W$ boson decay lepton asymmetry and the $W/Z$ cross section ratios in the PDF profiling analysis, we extract $\sin^2\theta^\ell_{\rm eff}=0.23154\pm0.00024$. The additional constraints reduce both the PDF uncertainty and the spread among modern PDF sets. The result is in excellent agreement with the 2026 SM indirect determination and is competitive with the most precise $e^+e^-$ collider measurements. The profiled PDFs also provide a useful input for reducing PDF uncertainties in other LHC extractions.


Further improvement is expected with the Run III data, corresponding to about 326 fb$^{-1}$ delivered by the LHC at 13.6 TeV.
For the combined CMS data samples at 13 and 13.6 TeV, we estimate that the total uncertainty in $\sin^2\theta^\ell_{\rm eff}$ can reach about 0.00020, with a PDF contribution of 0.00012. A combination with LHCb data at 13 TeV and the ATLAS 8 TeV analysis currently under study would provide an additional reduction in uncertainty.  This ushers a new era of precision electroweak measurement at the LHC.

Additional plots and tables, including the profiled PDF nuisance parameters and their correlation matrices for 19 NNLO PDF sets, are provided as Supplemental Material~\cite{PRD_supp} [URL will be inserted by publisher].

Research supported by the U.S. Department of Energy under University of Rochester grant number DE-SC0008475 and by  MSIP and NRF (Republic of Korea).
\bibliography{MixingAngle_PRD}
\clearpage
\appendix
\section{Supplementary Material: Effects of PDF profiling with CMS $A_4$, $W_{\rm asym}$ and $W/Z$ ratio measurements}
The Supplementary Material includes Table I which lists  the numerical values for the points shown in Fig.~\ref{Fig_2N} in the main paper, as well as additional plots illustrating the additional constraints on parton distribution functions.  The profiled PDF nuisance parameters are included as three tables for the CT18Znnlo, NNPDF4.0nnlo and PDF4LHC21nnlo. All the numbers in Table 1, and the complete list of eigenvalues and errors are included in a xlsx spreadsheet.

Specifically Table I shows  the values of $\sin^2\theta^\ell_{\rm eff}$ (in units of $10^{-5}$) before and after profiling with  the CMS 13 TeV  $A_4$ distribution (63 points) for 19 PDF sets.  
Also shown are the values extracted by including the CMS 13 TeV $W$-decay lepton asymmetry and the CMS W/Z cross section ratios (total of 83 points)  in the profiling. 
The  total uncertainties include contributions from statistical, experimental systematic, theoretical and PDF sources.  The column marked "Diff. from  \textsc{CT18Z}"  is the difference from  $\sin^2\theta^\ell_{\rm eff}$ extracted with the 
\textsc{CT18Znnlo} PDF set.
Note, the $\chi^2$ values before profiling do not include PDF errors, while the $\chi^2$ values after profiling include PDF errors.

The improved extraction is achieved by incorporating additional $W$ and $Z$ boson data, which constrain the parton distribution functions in the region around $Q \approx M_Z$ and $x=0.001$--$0.1$. 

\begin{figure*}
    \centering
    \includegraphics[width=2.1in,height=2.1in]{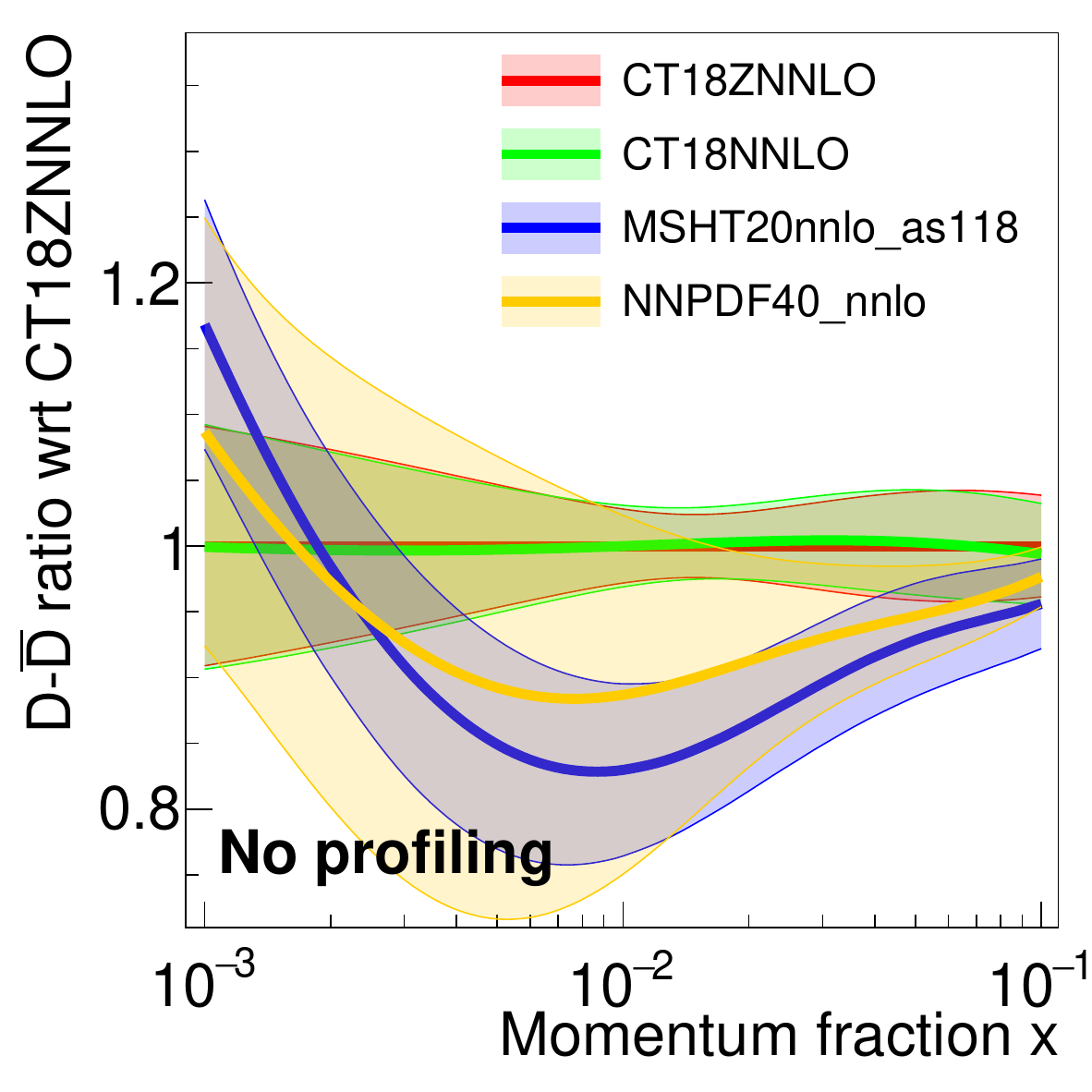}
    \includegraphics[width=2.1in,height=2.1in]{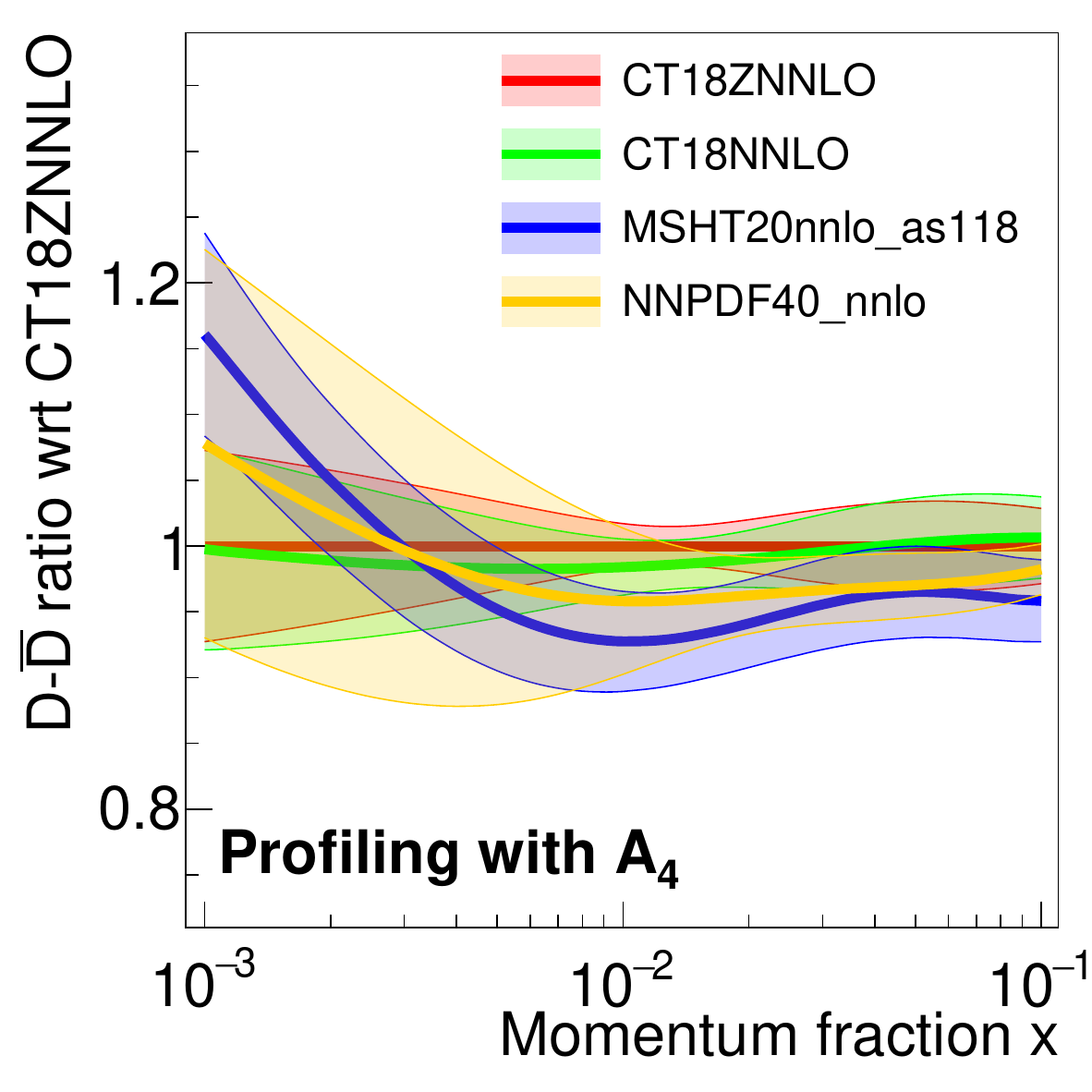}
    \includegraphics[width=2.1in,height=2.1in]{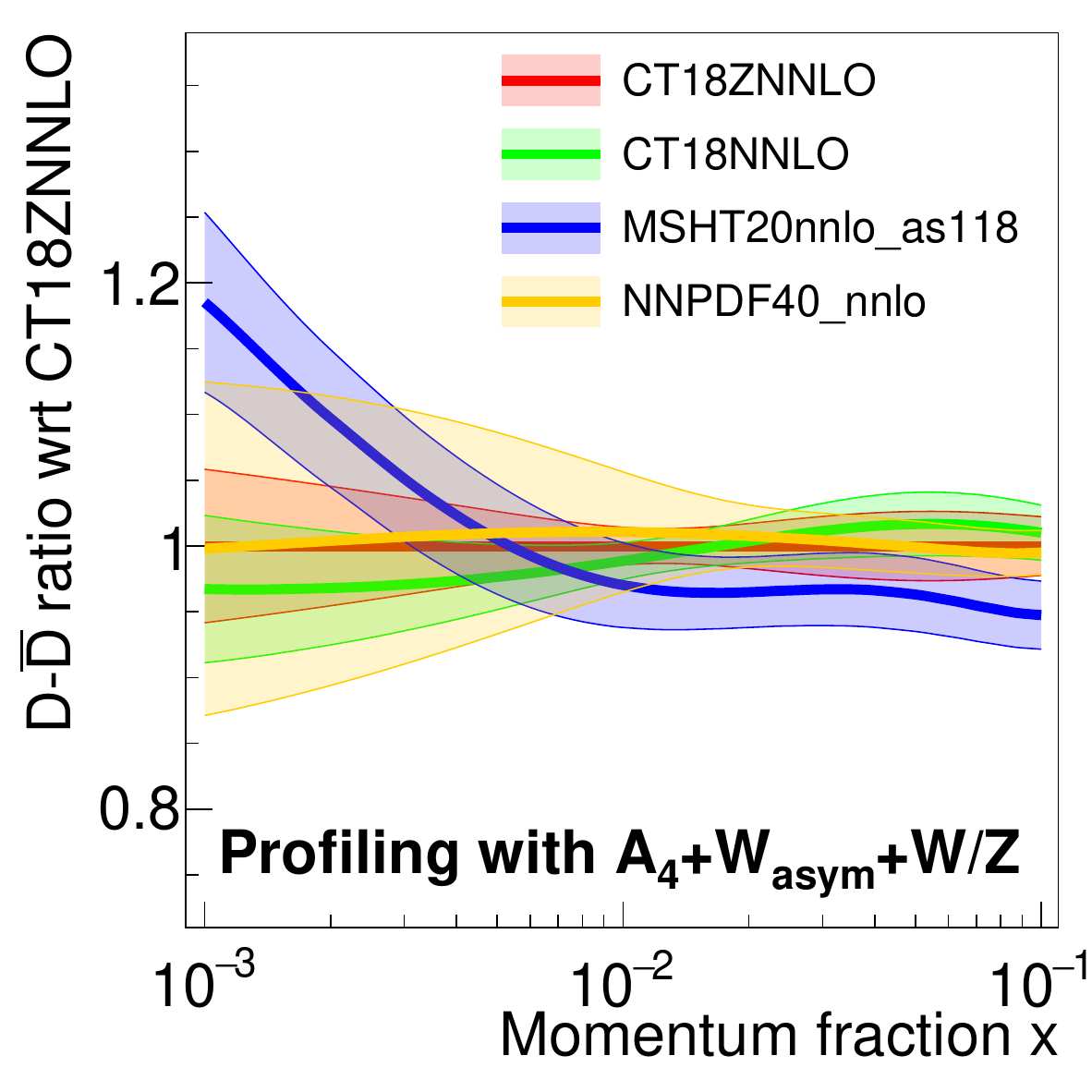}
    \caption{Ratios of the $D-\bar{D}$ distributions with respect to CT18Z for several PDF sets at $Q = M_Z$. The left panel shows the results without profiling, the middle panel includes profiling with $A_4$ measurement, and the right panel includes profiling with $A_4$, $W_{\rm asym}$, and $W/Z$ measurements.}
    \label{Fig_11}
\end{figure*}

\begin{figure*}
    \centering
    \includegraphics[width=2.1in,height=2.1in]{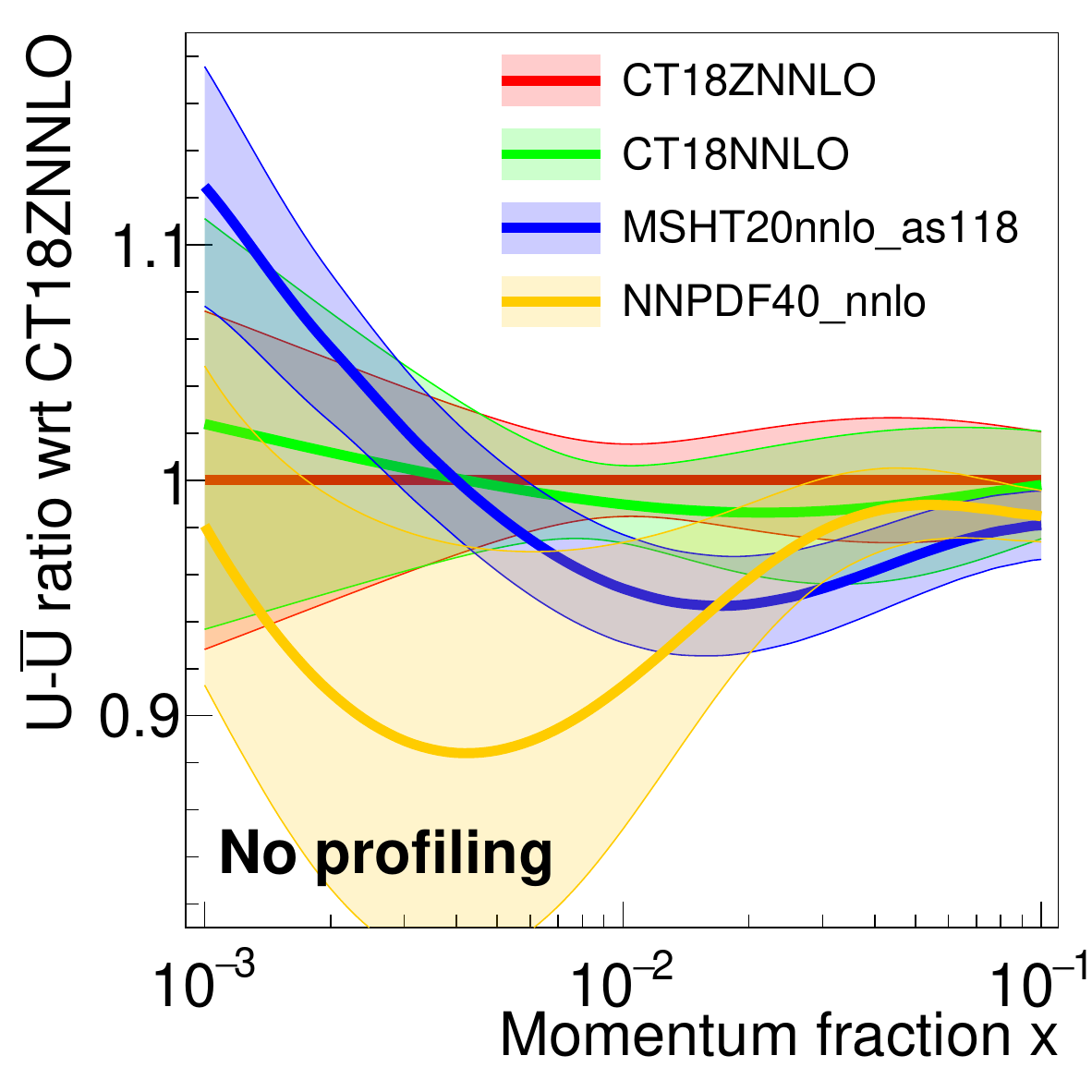}
    \includegraphics[width=2.1in,height=2.1in]{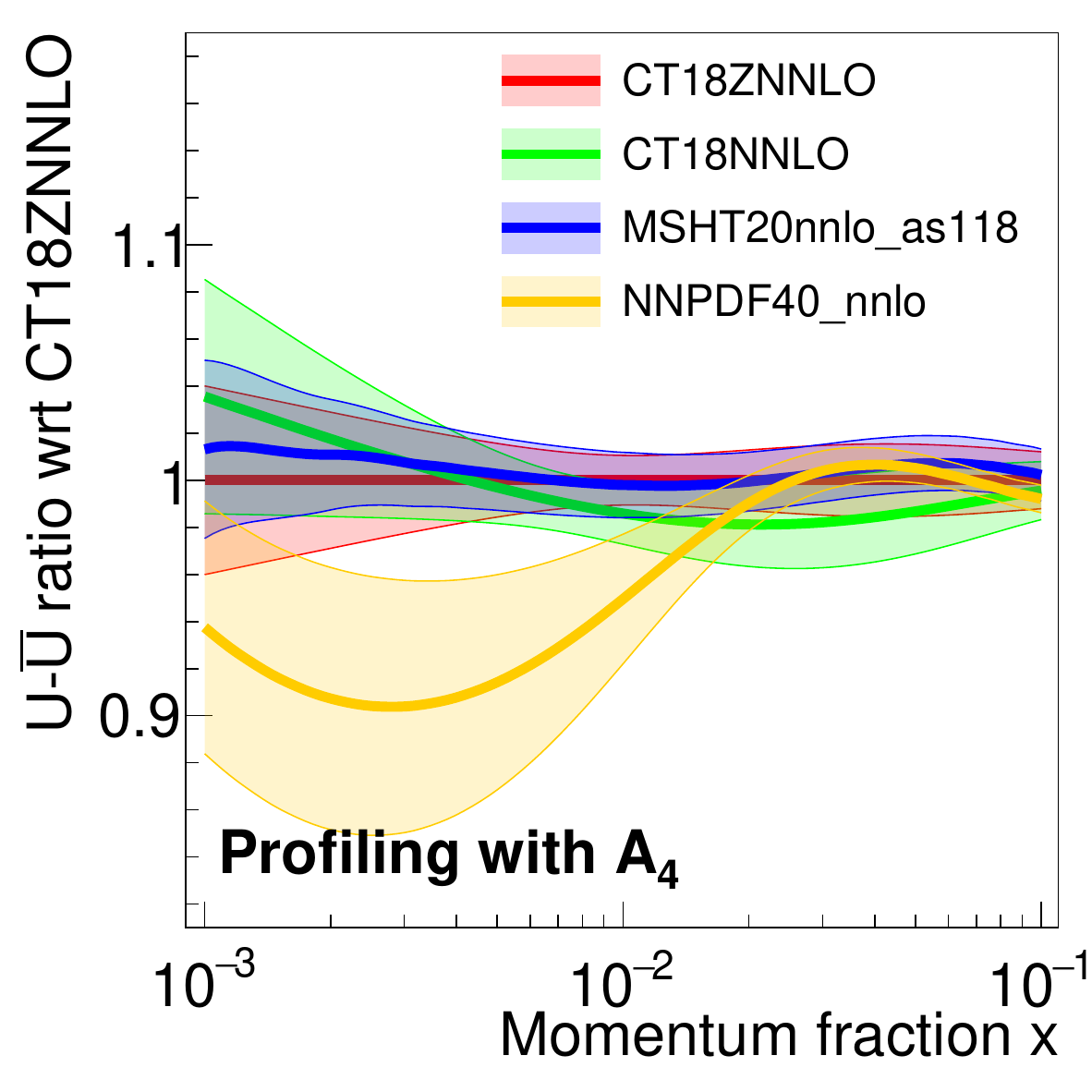}
    \includegraphics[width=2.1in,height=2.1in]{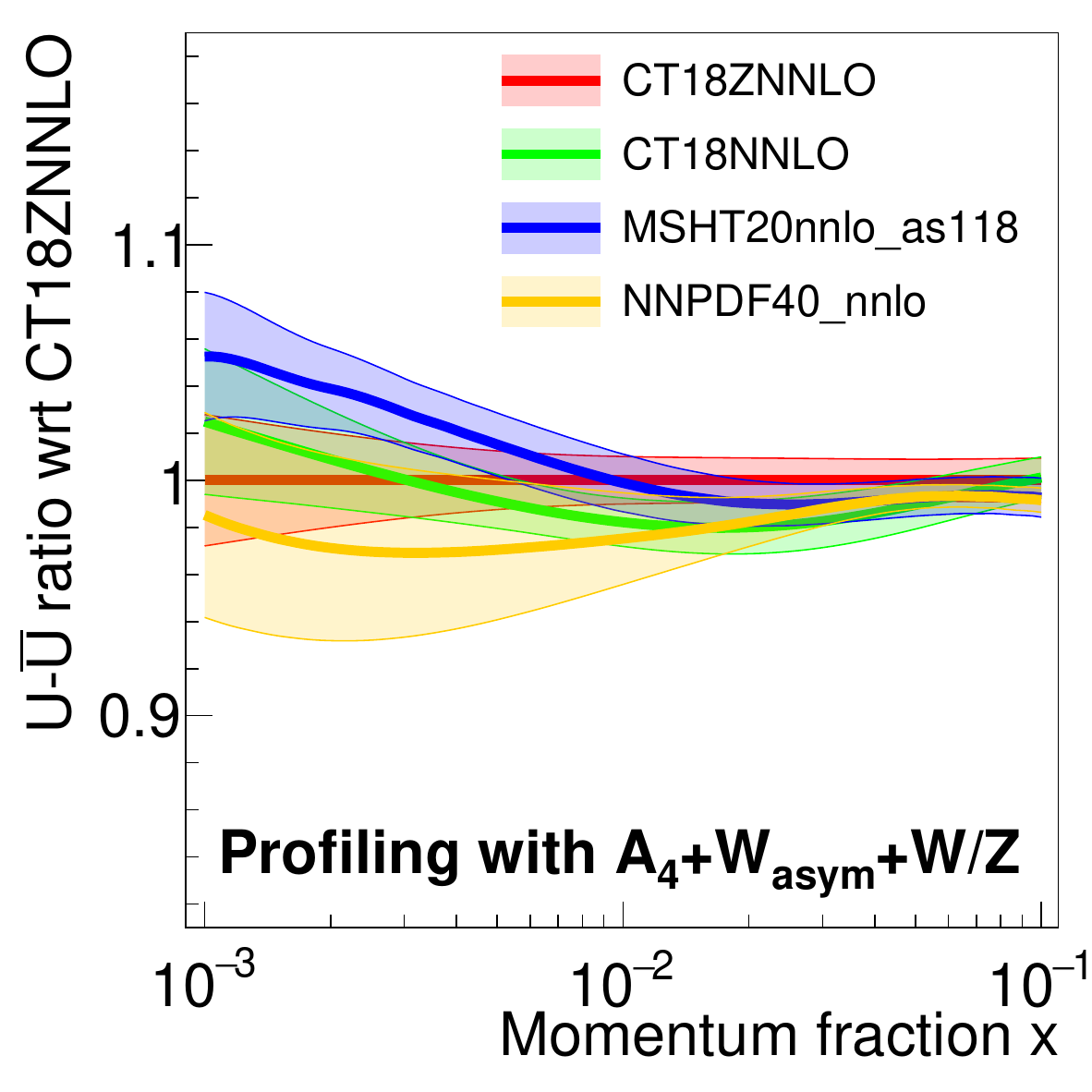}
    \caption{Ratios of the $U-\bar{U}$ distributions with respect to CT18Z for several PDF sets at $Q = M_Z$. The left panel shows the results without profiling, the middle panel includes profiling with $A_4$ measurement, and the right panel includes profiling with $A_4$, $W_{\rm asym}$, and $W/Z$ measurements.}
    \label{Fig_12}
\end{figure*}

The most relevant PDF quantities for the mixing angle extraction are the valence-quark distributions, defined as $d_v = d- \bar{d}$ and $u_v = u- \bar{u}$, as the mixing angle measurement in proton-proton collisions relies on the asymmetry in the  momentum distributions between quarks and antiquarks. Although the momentum difference mainly arises from up and down quarks, some PDF collaborations allow asymmetric distribution for other flavors as well. Therefore, for consistent comparison across different PDFs, it is useful to examine the $D-\bar{D}$ and $U-\bar{U}$ distributions, where $D=d+s+b$ and $U=u+c$. Figures \ref{Fig_11} and \ref{Fig_12} show how the $D-\bar{D}$ and $U-\bar{U}$ PDFs are constrained before and after profiling, illustrating how different PDFs converge after profiling with  additional $W$ and $Z$ boson measurements.

\begin{figure*}
    \centering
    \includegraphics[width=2.1in,height=2.1in]{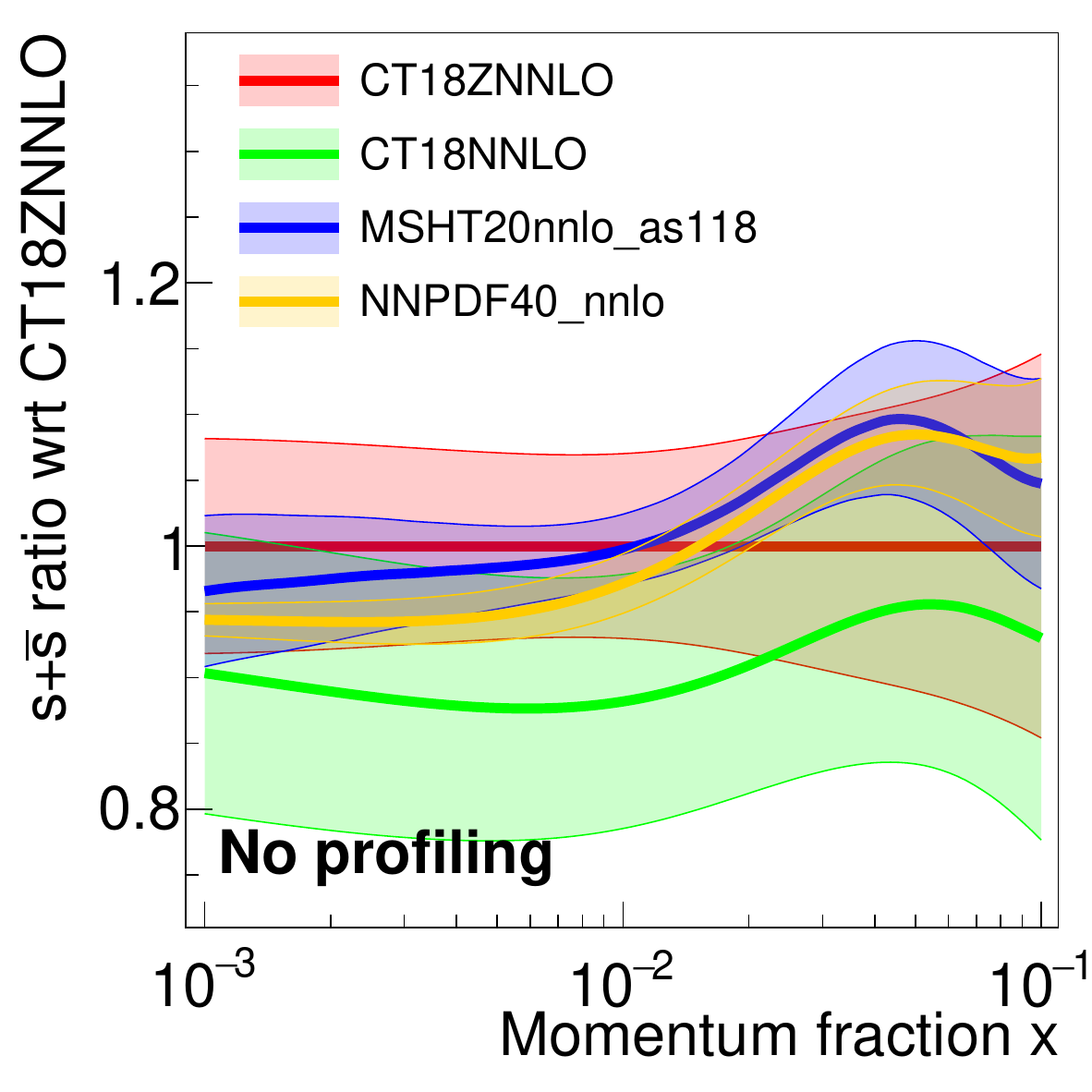}
    \includegraphics[width=2.1in,height=2.1in]{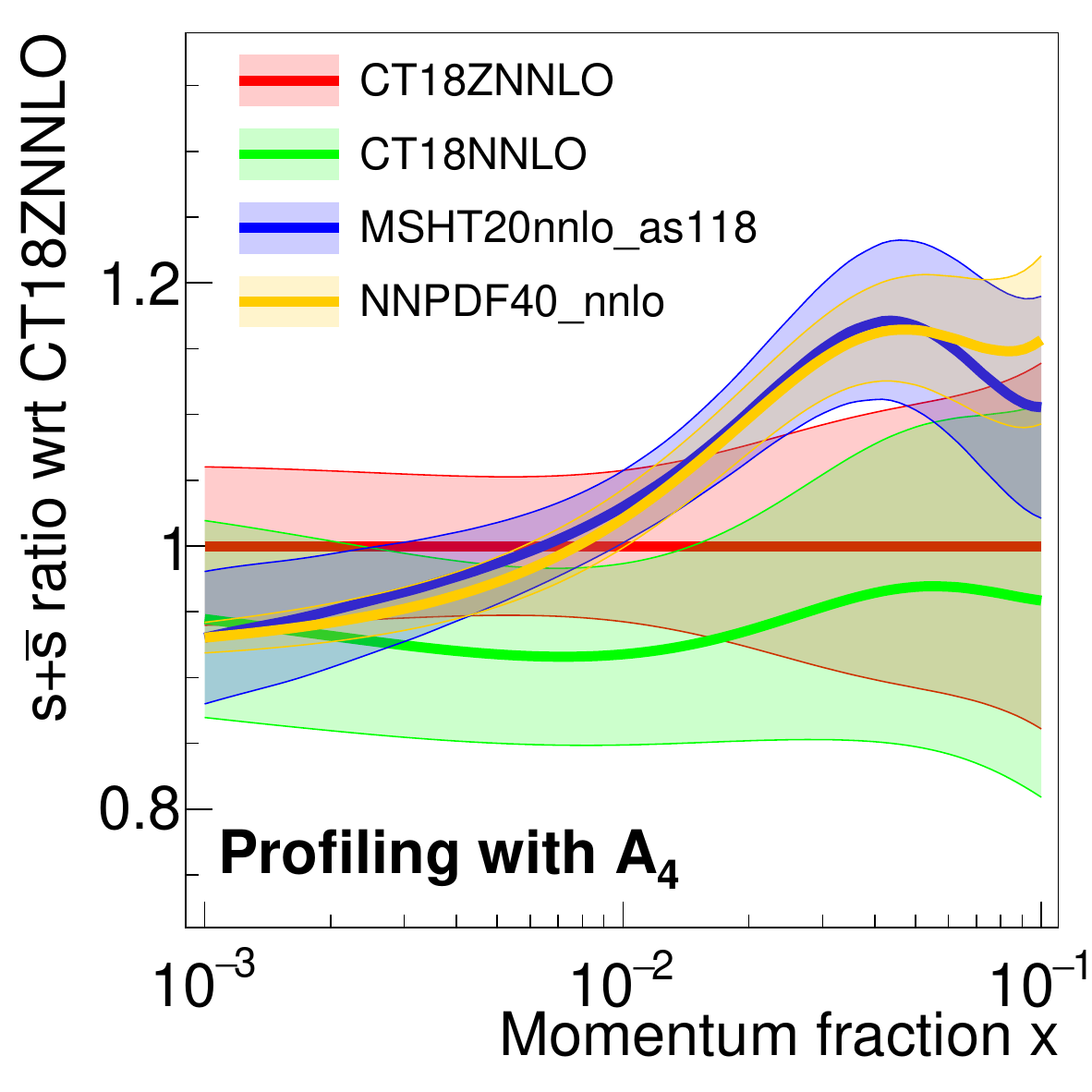}
    \includegraphics[width=2.1in,height=2.1in]{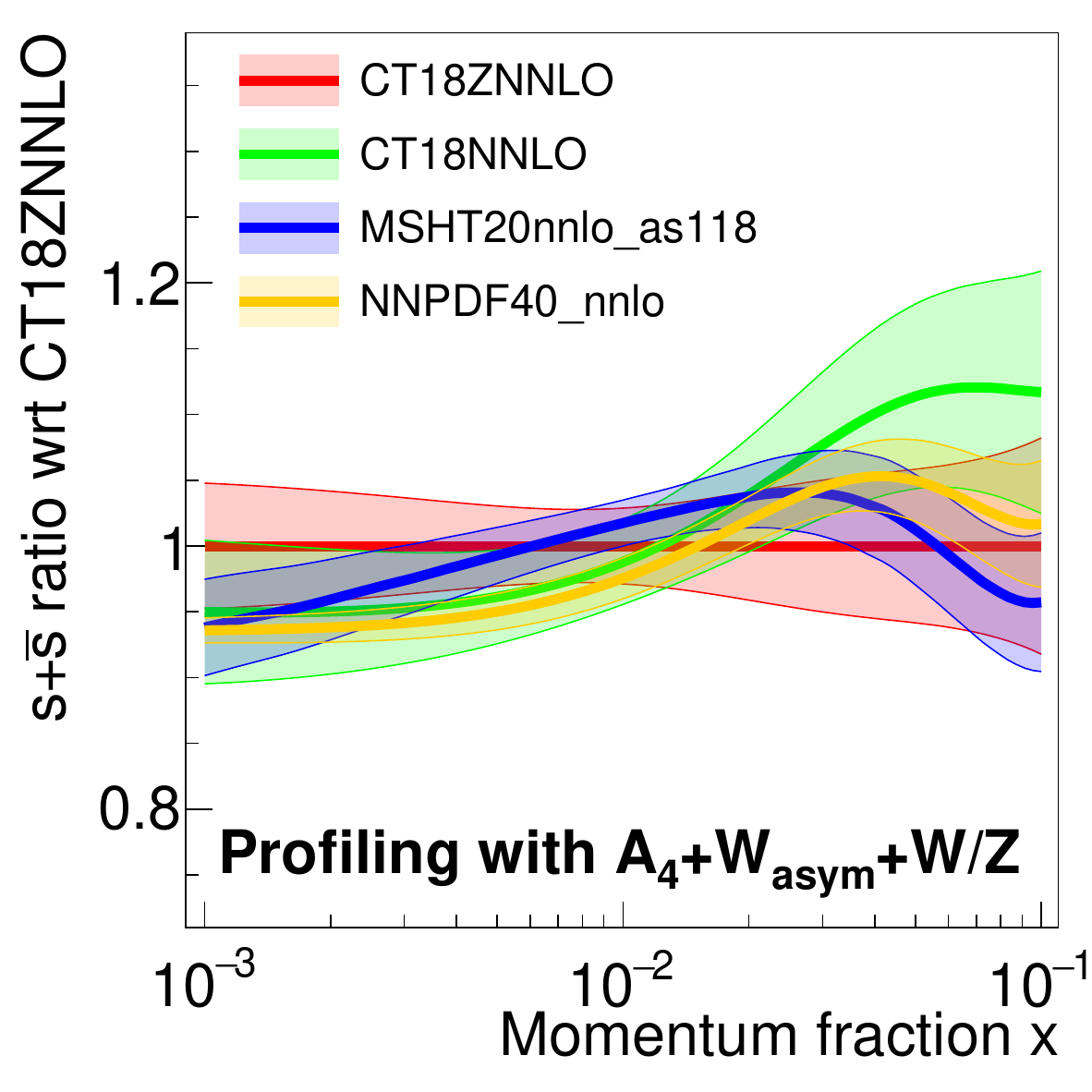}
    \caption{Ratios of the $s+\bar{s}$ distributions with respect to CT18Z for several PDF sets at $Q = M_Z$. The left panel shows the results without profiling, the middle panel includes profiling with $A_4$ measurement, and the right panel includes profiling with $A_4$, $W_{\rm asym}$, and $W/Z$ measurements.}
    \label{Fig_13}
\end{figure*}

It is also informative to examine the $s+\bar{s}$ PDF, which is primarily constrained by the $W/Z$ cross section ratio measurements, as shown in  Figure~\ref{Fig_13}. The CT18 PDF exhibits a smaller $s+\bar{s}$ component than the other PDF sets because the global PDF fit does not include some LHC data that constrain the strange quark at high $Q^2$.
This results in a higher mixing angle extracted with CT18. The inclusion of the $W/Z$ ratio measurement brings the $s+\bar{s}$ PDF in  CT18 to better agreement with  the other PDFs and results in a  value of the mixing angle which is closer to that obtained with CT18Z. The PDF nuisance parameter shifts after profiling are shown in Tables II, III and IV for  CT18ZNNLO, NNPDF40 and PDF4LHC21, respectively.

%
\begin{table*}[t]
\label{Table1}
\scriptsize
\setlength{\tabcolsep}{3pt}
\renewcommand{\arraystretch}{0.90}
\begin{center}																					\begin{tabular}{|c||c|c|c||c|c|c||c|c|c|} \hline
\multirow{4}{*}{PDF}    &\multicolumn{3}{c||}{$A_{4}$ without profiling} &\multicolumn{3}{c||}{$A_{4}$ profiling} &\multicolumn{3}{c|}{$A_{4}+W_{\rm asym}+W/Z$ profiling} \\
&\multicolumn{3}{c||}{(63 data points)} &\multicolumn{3}{c||}{(63 data points)} &\multicolumn{3}{c|}{(83 data points)} \\ \cline{2-10}
& \multirow{2}{*}{$\sin^2\theta^\ell_{\rm eff}$} & Diff. from & \multirow{2}{*}{$\chi^2$} & \multirow{2}{*}{$\sin^2\theta^\ell_{\rm eff}$} & Diff. from & \multirow{2}{*}{$\chi^2$} & \multirow{2}{*}{$\sin^2\theta^\ell_{\rm eff}$} & Diff. from & \multirow{2}{*}{$\chi^2$} \\
&  & \textsc{CT18Z} & & & \textsc{CT18Z} & & & \textsc{CT18Z} & \\ \hline\hline
NNPDF40 &                               &               &               &                               &               &               &                               &               &               \\
nnlo\_as\_01180\_hessian~\cite{NNPDF:2021njg}   &       23107 $\pm$ 49  &       -59     &       89      &       23130 $\pm$ 24  &       -22     &       66      &       23157 $\pm$ 23  &       3       &       94\\
nnlo\_as\_01180\_qed~\cite{NNPDF:2024djq}       &       23102 $\pm$ 46  &       -64     &       108     &       23144 $\pm$ 23  &       -8      &       69      &       23166 $\pm$ 22  &       12      &       97\\
nnlo\_as\_01180\_mhou~\cite{NNPDF:2024dpb}      &       23101 $\pm$ 45  &       -65     &       79      &       23110 $\pm$ 27  &       -42     &       64      &       23150 $\pm$ 24  &       -4      &       101\\
nnlo\_as\_01180\_qed\_mhou~\cite{NNPDF:2024djq} &       23103 $\pm$ 46  &       -63     &       93      &       23136 $\pm$ 24  &       -16     &       69      &       23164 $\pm$ 23  &       10      &       99\\
an3lo\_as\_01180~\cite{NNPDF:2024nan}   &       23129 $\pm$ 46  &       -37     &       72      &       23132 $\pm$ 25  &       -20     &       65      &       23158 $\pm$ 23  &       4       &       100\\
an3lo\_as\_01180\_qed~\cite{NNPDF:2024nan}      &       23129 $\pm$ 49  &       -37     &       78      &       23147 $\pm$ 23  &       -5      &       68      &       23160 $\pm$ 22  &       6       &       103\\
an3lo\_as\_01180\_mhou~\cite{NNPDF:2024nan}     &       23106 $\pm$ 47  &       -60     &       78      &       23120 $\pm$ 26  &       -32     &       65      &       23146 $\pm$ 25  &       -8      &       97\\
an3lo\_as\_01180\_qed\_mhou~\cite{NNPDF:2024nan}        &       23102 $\pm$ 44  &       -64     &       90      &       23137 $\pm$ 23  &       -15     &       69      &       23161 $\pm$ 23  &       7       &102\\ \hline
\textsc{CTEQ}   &                               &               &               &                               &               &               &                               &               &               \\
\textsc{CT18nnlo}~\cite{Hou:2019efy}    &       23204 $\pm$ 64  &       38      &       92      &       23166 $\pm$ 36  &       14      &       64      &       23167 $\pm$ 24  &       13      &       86\\
{\bf \textsc{CT18Znnlo}}~\cite{Hou:2019efy}     &       23166 $\pm$ 56  &       0       &       70      &       23152 $\pm$ 32  &       0       &       65      &       {\bf 23154 $\pm$ 24}    &       0       &83\\
\textsc{CT18Annlo}~\cite{Hou:2019efy}   &       23179 $\pm$ 54  &       13      &       76      &       23164 $\pm$ 28  &       12      &       67      &       23157 $\pm$ 23  &       3       &       82\\
\textsc{CT18Xnnlo}~\cite{Hou:2019efy}   &       23194 $\pm$ 53  &       28      &       86      &       23171 $\pm$ 30  &       19      &       64      &       23158 $\pm$ 24  &       4       &       88\\
\textsc{CT18qed-proton}~\cite{Xie:2023qbn}      &       23204 $\pm$ 64  &       38      &       93      &       23164 $\pm$ 36  &       12      &       64      &       23154 $\pm$ 24  &       0       &       82\\
\textsc{CT18As\_LatNNLO}~\cite{Hou:2022onq}     &       23177 $\pm$ 75  &       11      &       74      &       23108 $\pm$ 43  &       -44     &       64      &       23155 $\pm$ 35  &       1       &       84\\ \hline
MSHT20  &                               &               &               &                               &               &               &                               &               &               \\
MSHT20nnlo\_as118~\cite{Bailey:2020ooq} &       23053 $\pm$ 46  &       -113    &       200     &       23115 $\pm$ 30  &       -37     &       75      &       23119 $\pm$ 26  &       -35     &       106\\
MSHT20qed\_nnlo~\cite{Cridge:2021pxm}   &       23055 $\pm$ 59  &       -111    &       188     &       23124 $\pm$ 31  &       -28     &       73      &       23142 $\pm$ 27  &       -12     &       98\\
MSHT20an3lo\_as118~\cite{McGowan:2022nag}       &       23101 $\pm$ 47  &       -65     &       138     &       23138 $\pm$ 29  &       -14     &       69      &       23159 $\pm$ 26  &       5       &       85\\
MSHT20qed\_an3lo~\cite{Cridge:2023ryv}  &       23101 $\pm$ 52  &       -65     &       126     &       23137 $\pm$ 31  &       -15     &       69      &       23151 $\pm$ 28  &       -3      &       86\\ \hline
PDF4LHC21\_40~\cite{PDF4LHCWorkingGroup:2022cjn}        &       23122 $\pm$ 83  &       -44     &       80      &       23133 $\pm$ 33  &       -19     &       62      &       23139 $\pm$ 27  &       -15     &82\\ \hline
\end{tabular}
\caption{
Values of $\sin^2\theta^\ell_{\rm eff}$ (in units of $10^{-5}$) before and after profiling with  the CMS 13 TeV  $A_4$ distribution (63 points) for 19 PDF sets.  
Also shown are the values extracted by including the CMS 13 TeV $W$-decay lepton asymmetry and the CMS W/Z cross section ratios (total of 83 points)  in the profiling. 
The  total uncertainties include contribution from statistical, experimental systematic, theoretical and PDF sources.  The column marked "Diff. from  \textsc{CT18Z}"  is the difference from  $\sin^2\theta^\ell_{\rm eff}$ extracted with the 
\textsc{CT18Znnlo} PDF set.
Note, the $\chi^2$ values before profiling do not include PDF errors, while the $\chi^2$ values after profiling include PDF errors.
 }
\end{center}
\vspace{-0.1in}
\end{table*}


\begin{table}[!htb]
\label{Table2}
\centering
\scriptsize
\setlength{\tabcolsep}{2pt}
\setlength{\abovecaptionskip}{2pt}
\setlength{\belowcaptionskip}{2pt}
\renewcommand{\arraystretch}{0.64}
\caption{Shifts and uncertainties of the 29 nuisance parameters corresponding to the CT18Z set after profiling with $A_4$, $W_{\rm asym}$, and $W/Z$ fiducial cross-section ratio. The prior uncertainties are defined to be 1.0, and before profiling the shifts are zero and the errors are  1.0  by definition.}
\resizebox{\columnwidth}{!}{\begin{tabular}{lcc}
\toprule
Index & Shift & Uncertainty \\
\midrule
1 & -0.2698 & 0.8694 \\
2 & 0.4215 & 0.8383 \\
3 & 0.7903 & 0.9111 \\
4 & 0.1190 & 0.9624 \\
5 & 0.3964 & 0.9353 \\
6 & -0.1410 & 0.9035 \\
7 & 0.1101 & 0.8725 \\
8 & 0.1668 & 0.8608 \\
9 & -0.3094 & 0.9057 \\
10 & 0.4435 & 0.8992 \\
11 & 0.3144 & 0.8146 \\
12 & -0.2324 & 0.8963 \\
13 & 0.3711 & 0.8928 \\
14 & 0.1572 & 0.8638 \\
15 & -0.5092 & 0.9180 \\
\bottomrule
\end{tabular}

\hspace{1cm}
\begin{tabular}{lcc}
\toprule
Index & Shift & Uncertainty \\
\midrule
16 & -0.0998 & 0.8569 \\
17 & 0.4871 & 0.8999 \\
18 & -0.5412 & 0.9113 \\
19 & -0.1664 & 0.7318 \\
20 & -0.1035 & 0.9096 \\
21 & 0.2783 & 0.9340 \\
22 & 0.3516 & 0.9429 \\
23 & -0.6438 & 0.8087 \\
24 & 0.1258 & 0.8168 \\
25 & 0.1400 & 0.8849 \\
26 & -0.1593 & 0.8616 \\
27 & -0.4221 & 0.9555 \\
28 & 0.0709 & 0.8960 \\
29 & -0.8794 & 0.8651 \\
- & - & - \\
\bottomrule
\end{tabular}}
\end{table}

\begin{table}[!htb]
\label{Table3}
\centering
\scriptsize
\setlength{\tabcolsep}{2pt}
\setlength{\abovecaptionskip}{2pt}
\setlength{\belowcaptionskip}{2pt}
\renewcommand{\arraystretch}{0.64}
\caption{Shifts and uncertainties of the 50 nuisance parameters corresponding to the NNPDF40 set after profiling with $A_4$, $W_{\rm asym}$, and $W/Z$ fiducial cross-section ratio. The prior uncertainties are defined to be 1.0, and before profiling the shifts are zero and the errors are 1.0 by definition.}
\resizebox{\columnwidth}{!}{\begin{tabular}{lcc}
\toprule
Index & Shift & Uncertainty \\
\midrule
1 & -0.4062 & 0.8326 \\
2 & -0.4298 & 0.9169 \\
3 & 0.0805 & 0.8975 \\
4 & -0.1186 & 0.9094 \\
5 & -0.7067 & 0.7574 \\
6 & 0.1617 & 0.8375 \\
7 & -0.0801 & 0.9048 \\
8 & -1.0562 & 0.5966 \\
9 & -0.1790 & 0.9244 \\
10 & -0.5657 & 0.9231 \\
11 & -0.1075 & 0.8061 \\
12 & -0.2746 & 0.8974 \\
13 & 0.0388 & 0.6245 \\
14 & -0.1622 & 0.8423 \\
15 & -0.4135 & 0.9009 \\
16 & 0.1044 & 0.8519 \\
17 & 0.2488 & 0.9040 \\
18 & 0.2466 & 0.9482 \\
19 & -0.9258 & 0.7836 \\
20 & 0.6645 & 0.9102 \\
21 & -0.7575 & 0.8875 \\
22 & 0.3261 & 0.9304 \\
23 & 0.4327 & 0.8657 \\
24 & 0.2389 & 0.8527 \\
25 & 0.2369 & 0.8797 \\
\bottomrule
\end{tabular}

\hspace{1cm}
\begin{tabular}{lcc}
\toprule
Index & Shift & Uncertainty \\
\midrule
26 & 0.5480 & 0.8973 \\
27 & 0.1383 & 0.9270 \\
28 & 0.8098 & 0.9443 \\
29 & 0.0235 & 0.9041 \\
30 & 0.3823 & 0.9299 \\
31 & 0.0165 & 0.9051 \\
32 & 0.3545 & 0.9471 \\
33 & 0.2801 & 0.9055 \\
34 & -0.0945 & 0.8984 \\
35 & 0.3153 & 0.8991 \\
36 & -0.2786 & 0.8777 \\
37 & -0.0430 & 0.8761 \\
38 & 0.0634 & 0.9188 \\
39 & 0.1318 & 0.8710 \\
40 & 0.7378 & 0.9281 \\
41 & 0.1706 & 0.9480 \\
42 & -0.2648 & 0.9249 \\
43 & -1.2336 & 0.9398 \\
44 & 1.0300 & 0.9193 \\
45 & 0.1933 & 0.9446 \\
46 & -0.4014 & 0.9027 \\
47 & 0.5288 & 0.8923 \\
48 & -0.1129 & 0.9340 \\
49 & -0.3289 & 0.8625 \\
50 & 0.8782 & 0.9352 \\
\bottomrule
\end{tabular}}
\end{table}

\begin{table}[!htb]
\label{Table4}
\centering
\scriptsize
\setlength{\tabcolsep}{2pt}
\setlength{\abovecaptionskip}{2pt}
\setlength{\belowcaptionskip}{2pt}
\renewcommand{\arraystretch}{0.64}
\caption{Shifts and uncertainties of the 40 nuisance parameters corresponding to the PDF4LHC21 set after profiling with $A_4$, $W_{\rm asym}$, and $W/Z$ fiducial cross-section ratio. The prior uncertainties are defined to be 1.0, and before profiling the shifts are zero and the errors are  1.0  by definition.}
\resizebox{\columnwidth}{!}{\begin{tabular}{lcc}
\toprule
Index & Shift & Uncertainty \\
\midrule
1 & 0.4872 & 0.8003 \\
2 & -0.6364 & 0.5057 \\
3 & -0.5334 & 0.8277 \\
4 & 0.1796 & 0.6523 \\
5 & -0.1220 & 0.8241 \\
6 & 0.1208 & 0.8453 \\
7 & -0.4243 & 0.8410 \\
8 & 0.0296 & 0.7673 \\
9 & 0.9090 & 0.8440 \\
10 & -0.0278 & 0.9035 \\
11 & 0.8183 & 0.8017 \\
12 & -0.2330 & 0.8307 \\
13 & -0.3762 & 0.7775 \\
14 & 0.1165 & 0.8794 \\
15 & 0.0528 & 0.7779 \\
16 & -0.3485 & 0.7366 \\
17 & -0.5867 & 0.7208 \\
18 & -0.2765 & 0.7205 \\
19 & 0.3287 & 0.8237 \\
20 & 0.6397 & 0.8404 \\
\bottomrule
\end{tabular}

\hspace{1cm}
\begin{tabular}{lcc}
\toprule
Index & Shift & Uncertainty \\
\midrule
21 & 0.2801 & 0.7875 \\
22 & 0.1707 & 0.8620 \\
23 & 0.2645 & 0.8634 \\
24 & -0.0407 & 0.9062 \\
25 & -0.0920 & 0.7255 \\
26 & -0.3067 & 0.8405 \\
27 & 0.9956 & 0.8177 \\
28 & -0.4254 & 0.8364 \\
29 & 0.0389 & 0.7752 \\
30 & 0.0683 & 0.7927 \\
31 & -0.1432 & 0.8568 \\
32 & -0.6806 & 0.8588 \\
33 & 0.5062 & 0.7235 \\
34 & 0.6989 & 0.8568 \\
35 & 0.9569 & 0.8524 \\
36 & 0.0445 & 0.8049 \\
37 & 0.2833 & 0.7109 \\
38 & -0.8043 & 0.7568 \\
39 & -0.4103 & 0.6464 \\
40 & 1.4596 & 0.7171 \\
\bottomrule
\end{tabular}}
\end{table}
\end{document}